\documentstyle[12pt]{article}
\newcommand{\be}{\begin{equation}}
\newcommand{\ee}{\end{equation}}
\newcommand{\bea}{\begin{eqnarray}}
\newcommand{\eea}{\end{eqnarray}}
\newcommand{\ba}{\begin{array}}
\newcommand{\ea}{\end{array}}
\newcommand{\bmat}{\left(\ba}
\newcommand{\emat}{\ea\right)}

\newcommand{\norsl}{\normalsize\sl}
\newcommand{\norsc}{\normalsize\sc}
\begin{document}
%-------------------- Title page ----------------------------------
\begin{titlepage}

\title{
Nonlinear Realization of Partially Broken $N=2$ \\
Superconformal Symmetry in Four Dimensions   
}

\author{ \\
\norsc  Yoshinori GOTOH  \\
\norsl  Graduate School of Human and \\
\norsl  Environmental Studies, Kyoto University\\
\norsl  Kyoto 606-01, JAPAN\\
\\
and
\\
\\
\norsc       Tsuneo UEMATSU\thanks{Supported in part by
          the Monbusho Grant-in-Aid for Scientific Research
          No. C-09640345 and for Scientific Research on Priority Areas
          No. 09246215 } \\
\norsl  Department of Fundamental Sciences\\
\norsl  FIHS, Kyoto University\\
\norsl  Kyoto 606-01, JAPAN\\
\\
}

\date{}

\maketitle

\begin{abstract}
{\normalsize
We investigate the nonlinear realization of spontaneously broken $N=2$ 
superconformal symmetry in 4 dimensions. We particularly study Nambu-Goldstone
degrees of freedom for the partial breaking of $N=2$ superconformal symmetry 
down to $N=1$ super-Poincar{\'e} symmetry, where we get the chiral NG 
multiplet of dilaton and the vector NG multiplet of NG fermion of broken 
$Q$-supersymmetry. Evaluating the covariant differentials and supervielbeins
for the chiral as well as the full superspace, we obtain the nonlinear 
effective lagrangians.

}
\end{abstract}

\begin{picture}(5,2)(-310,-660)
\put(35,-120){KUCP-108}
\put(35,-135){July 1997}
\end{picture}

\vspace{2cm}
%\leftline{\hspace{1cm}hep-ph/9603338}
 
\thispagestyle{empty}
\end{titlepage}
\setcounter{page}{1}
\baselineskip 20pt
%\baselineskip 24pt

%----------------------- Text -----------------------------------
In the last few years there has been much interest in $N=2$ supersymmetric
gauge theory in the context of duality which provides us with the 
understanding of non-perturbative properties of supersymmetric field theories 
as well as string theories.  

Hughes, Liu and Polchinski \cite{HLP} pointed out that $N=2$ global 
supersymmetry can be partially broken down to $N=1$ supersymmetry by giving 
the argument on the way how to evade the existing no-go theorem of partial 
breaking of extended supersymmetry based on the supersymmetry current algebra.
They also explicitly constructed the four-dimensional supermembrane solution 
of the six-dimensional supersymmetric gauge theory, in which the second 
supersymmetry, in the equivalent four-dimensional $N=2$ theory, is 
spontaneously broken and the partial breaking is realized. While, Antoniadis, 
Partouche and Taylor \cite{APT} introduced the electric and magnetic 
Fayet-Iliopoulos terms  in the $N=2$ gauge theory of abelian vector multiplet 
and have shown that there occurs spontaneous breaking of $N=2$ to $N=1$ 
supersymmetry. This partial breaking induced by the Fayet-Iliopoulos terms 
has also been obtained by taking the flat limit of the $N=2$ supergravity 
theories \cite{FGP}.

Bagger and Galperin \cite{BG1,BG2} have studied the nonlinear realization of 
$N=2$ supersymmetry partially broken down to $N=1$ supersymmetry \cite{BW,SW}
and obtained the Nambu-Goldstone multiplet both for the cases; chiral
multiplet \cite{BG1} and vector-multiplet \cite{BG2}, and discussed 
the nonlinear transformation laws as well as the low-energy effective 
lagrangians \cite{B}.

Here in this paper, we shall investigate the nonlinear realization of
$N=2$ extended superconformal symmetry in four-dimensions, which is realized
for the case of vanishing $\beta$-function in the N=2 supersymmetric QCD.
We study the spontaneously breaking of this symmetry down to $N=1$ 
super-Poincar{\'e} symmetry. We identify the Nambu-Goldstone degrees of 
freedom corresponding to the broken charges. 
It turns out that there appear a vector multiplet  
of NG fermion of the broken second $Q$-supersymmetry as well as  a chiral 
multiplet of dilaton, axion and dilatino which are associated with the 
broken dilatation, chiral $U(1)$ rotation and the first $S$-supersymmetry 
generators. We note that the NG fermion of the broken second $S$-supersymmetry
can be expressed as the derivative of the true NG-fermion of the second 
$Q$-supersymmetry. We obtain the effective interaction for the
system of the dilaton multiplet coupled to the NG-vector multiplet.
 
Let us suppose a certain symmetry, characterized by a group $G$, which is
spontaneously broken down to its subsymmetry given by a subgroup
$H$. In such a case there appear Nambu-Goldstone (NG) fields that transform
nonlinearly under $G$ as the coordinates of the coset space $G/H$. In the
framework of the nonlinear realization, we can construct low-energy
effective lagrangians describing the interactions of massless NG particles.
Based on the framework \cite{V,O,UZ} for the nonlinear realization of space-time
symmetries, which is the modification of those for the internal symmetry
\cite{CWZ} we can investigate the spontaneously broken  extended supersymmetry
\cite{Fayet}.

We now consider the 4-dimensional N=2 superconformal group usually denoted
by $SU(2,2/2)$, the generators of which are those of conformal group: 
translation $P_\mu$, Lorentz rotation $M_{\mu\nu}$, conformal boost $K_\mu$,
and dilatation $D$ operators; together with $Q$-supersymmetry generators:
$Q_{\alpha A}$, $\bar{Q}_{\dot\alpha}^A$ ($A=1,2$);  
$S$-supersymmetry generators: $S^{\alpha A}$, $\bar{S}^{\dot\alpha}_A$ 
($A=1,2$) and $SU(2)$ generators, ${T_A}^B$, the chiral $U(1)_R$ charge 
$A$, and the $U(2)$ charge${B_A}^B \equiv {T_A}^B+\frac{1}{6}\delta_A^B A$. 
Some relevant commutation relations are the following:
\bea
&&[P_{\mu},K_{\nu}]=2i(g_{\mu \nu}D-M_{\mu \nu})\ , \
\{Q_{\alpha A},\bar{Q}_{\dot{\alpha}}^B\}=2\delta_A^B 
\sigma _{\mu \alpha \dot{\alpha}}P^{\mu}\ , \  
\{S_{\alpha}^A,\bar{S}_{\dot{\alpha} B}\}=2\delta^A_B
\sigma _{\mu \alpha \dot{\alpha}}K^{\mu}\nonumber\\
&&[Q_{\alpha A},D]=\frac{i}{2}Q_{\alpha A}\ \ ,\ \
[Q_{\alpha A},A]=\frac{3}{2}Q_{\alpha A}\ \ ,\ \
[\bar{S}_A^{\dot\alpha},D]=-\frac{i}{2}\bar{S}_A^{\dot\alpha} \ \ ,\ \ 
[S^{\alpha A},A]=-\frac{3}{2}S^{\alpha A}
\nonumber\\
&&\{Q_{\alpha A},S^{\beta B}\}=\delta_A^B [(\sigma^{\mu \nu})
_{\alpha}^{\ \beta}
M_{\mu \nu}-2i\delta_{\alpha}^{\beta}D]-4\delta_{\alpha}^{\beta}B_A^{\ B}
\nonumber\\
&&[Q_{\alpha A},K_{\mu}]=\sigma _{\mu \alpha \dot{\alpha}}
\bar{S}^{\dot{\alpha}}_A\ \ , \ \
[\bar{S}_A^{\dot{\alpha}},P_{\mu}]=\bar{\sigma}
_{\mu}^{\dot{\alpha} \alpha}Q_{\alpha A}\\
&& [Q_{\alpha A},T_B^{\ C}]=\delta_A^C Q_{\alpha B}-\frac{1}{2}
\delta_B^C Q_{\alpha A}\ \ , \ \
[\bar{S}^{\dot{\alpha}}_{A},T_B^{\ C}]=\delta_A^C 
\bar{S}^{\dot{\alpha}}_{B}-\frac{1}{2}
\delta_B^C \bar{S}^{\dot{\alpha}}_{A}\nonumber\\
&&[T_A^{\ B},T_C^{\ D}]=\delta_A^D T_C^{\ B}-\delta_C^BT_A^{\ D}
\qquad (A,B,C,D=1,2)
\nonumber
\eea

The various breaking patterns of this symmetry are illustrated in Fig.1. 
The spontaneous breaking of the pattern E has been the subject discussed 
in the literatures \cite{HLP,APT,FGP,BG1,BG2}. Here in this paper, we shall 
investigate the breaking pattern B. 

Now we note that the charge commutation relation between an unbroken charge
and a broken charge gives another broken charge. This relation 
%\be
%\left[\  \mbox{unbroken charge , broken charge}\ \right]=\mbox{broken charge}
%\label{eqn:a1}
%\ee
puts the constraints on NG particles.
For example, when conformal group is spontaneously broken to Poincar{\'e} 
group, the NG field $\phi_{\mu}$ associated with the conformal boost $K_\mu$ 
is not an independent NG degree of freedom, but is related to the dilaton 
$\sigma$, the NG particle for dilatation, as 
$\phi_{\mu}\sim \partial_{\mu}\sigma$. 
This relation is obtained through the use of one of the commutation relations 
of conformal algebra 
\be
[P_{\mu},K_{\nu}]=2i(g_{\mu \nu}D-M_{\mu \nu})
\label{eqn:a2}
\ee
and the Jacobi identity for $P_{\mu},K_{\nu}$ and $\sigma$ 
\be
[[P_{\mu},K_{\nu}],\sigma]+[[K_{\nu},\sigma],P_{\mu}]+[[\sigma,P_{\mu}],
K_{\nu}]=0 .
\label{eqn:a3}
\ee 
The vacuum expectation value of (\ref{eqn:a3}) gives 
$\langle 0|[K_{\nu} , \partial_{\mu}\sigma]|0\rangle
=-2g_{\mu \nu}\langle 0|[D , \sigma]|0\rangle \neq 0$. 

First we consider the case where $D$=4 $N$=2 superconformal group is 
spontaneously broken down to $N$=1 superconformal group, i.e. the breaking
A in Fig.1. In this case, we have the broken generators 
$S^{\alpha 2}({\bar S}_2^{\dot\alpha}$), 
$Q_{\alpha 2}({\bar Q}_{\dot\alpha})$, $T$ and $\bar{T}$, while
$B_1^1=3R/4$ and $B_2^2$ remain unbroken. Here $R$ is the $U(1)_R$ for
N=1 superconformal group. 

The constraints on NG particles are obtained in the same manner. 
The following commutation relations: 
\be
[\bar{S}_2^{\dot{\alpha}},P_{\mu}]=\bar{\sigma}
_{\mu}^{\dot{\alpha} \alpha}Q_{\alpha 2}\ ,\ \ [Q_{\alpha 1},T]=Q_{\alpha 2}\ 
,\ 
\{Q_{\alpha 1},S^{\beta 2}\}=-4\delta_{\alpha}^{\beta}\bar{T}
\label{eqn:a5}
\ee 
give rise to the constraints on $N$=1 NG multiplets 
$\chi^{\alpha}(x,\theta,\bar{\theta}),\ 
v(x,\theta,\bar{\theta})$ and $\psi_{\alpha 2}(x,\theta,\bar{\theta})$ 
%(dimension is $-\frac{1}{2},\ 0$ and $\frac{1}{2}$) 
for $Q_{\alpha 2},\ T$ and $S^{\beta 2}$. 
%(dimension is $\frac{1}{2},\ 0$ and $-\frac{1}{2}$).
The Jacobi identities lead to
\be
\bar{\psi}^{2{\dot\alpha}}\sim \bar{\sigma}_{\mu}^{{\dot\alpha}\alpha}
\partial^{\mu}\chi_\alpha\ ,\ v\sim \{Q_{1}^\alpha,\chi_\alpha \}\ ,
\ \bar{\psi}^{2{\dot\alpha}}\sim  [\bar{Q}_1^{\dot\alpha},v]
%\sim \bar{\sigma}_{\mu}\partial^{\mu}\chi.
\label{eqn:a6}
\ee 
These transformations (\ref{eqn:a6}) show that there is an independent 
$N$=1 NG multiplet $\chi^{\alpha}(x,\theta,\bar{\theta})$ which contains 
component fields $\chi^{\alpha},\ v$ and $\psi_{\alpha 2},$
moreover we can identify these fields with Abelian gaugino, D-term and 
the derivative of gaugino, respectively. Now we impose the chirality 
condition on $\chi^{\alpha}(x,\theta,\bar{\theta})$.
Namely, $\chi^{\alpha}(x,\theta,\bar{\theta})$
is a vector multiplet with spin $\frac{1}{2}$ and 1 fields.
The spin 1 field does not really correspond to a NG degree of freedom, but a 
superpartner of the NG fermion. For this reason, we may call this multiplet
as NG-Maxwell multiplet.

From the argument based on the charge commutation relation 
(\ref{eqn:a5}), we expect
that if $Q$-supersymmetry is spontaneously broken, then the
corresponding $S$-supersymmetry is broken as well and the NG fermion 
corresponding to the broken $S$-supercharge is written as a derivative of the 
NG fermion corresponding to the broken $Q$-supercharge. We shall 
present a proof for this statement 
by using a spectral representation \cite{Hig}. Here we use,
for simplicity, the four-component Majorana fermion representation.

Let us introduce $Q$-supercurrent $J_{\mu \alpha}$, $S$-supercurrent 
$S_{\mu \alpha}$ and consider the spectral representation of the 
two-point function. According to the $PCT$ invariance, Lorentz covariance, 
the spectral condition as well as parity invariance, we can write down the 
two-point function as 
\bea
\langle 0|\{J_{\mu \alpha}(x),\bar{\chi}_{\beta}(y)\}|0 \rangle &=&
\int_0^{\infty}dm^2
[\rho_1(m^2)i\partial_{\mu}\delta_{\alpha \beta}+\rho_2(m^2)(\gamma_
{\mu})_ {\alpha \beta}+\rho_3(m^2)i\partial^{\nu}(\sigma_{\mu \nu})_
{\alpha \beta}]\nonumber\\
&&\times i\Delta(x-y;m^2).
\label{eqn:a9}
\eea
where $\rho_i(m^2)$ ($i=1,2,3$) are the spectral functions.
$Q$-supercurrent conservation $\partial^{\mu}J_{\mu \alpha}=0$ 
as well as $S$-supercurrent conservation 
$(\gamma^{\mu})_{\alpha \beta}J_{\mu \beta}=0$ 
lead to
\be
m^2\rho_1(m^2)=0\ ,\ \rho_2(m^2)=0, \ \rho_3(m^2)=-\frac{1}{3}\rho_1(m^2)
\label{eqn:a11}
\ee 
Hence, we can write down the spectral representation of the 
two-point function as ($\rho_1(m^2) \equiv \rho(m^2)$):
\be
\langle 0|\{J_{\mu \alpha}(x),\bar{\chi}_{\beta}(y)\}|0\rangle =
\int_0^{\infty}dm^2
\rho(m^2)[i\partial_{\mu}\delta_{\alpha \beta}
-\frac{1}{3}i\partial^{\nu}(\sigma_{\mu \nu})_{\alpha \beta}]
\ i\Delta(x-y;m^2),\label{eqn:a14}
\ee
with $m^2\rho(m^2)=0$, which implies 
\be
\rho(m^2)=c\delta (m^2).
\label{eqn:a16}
\ee 
We note that if  $c\neq 0$,  $Q$-SUSY is spontaneously broken and 
the corresponding NG fermion is $\chi_{\alpha}$.

Next we calculate the following vacuum expectation value of 
the equal-time commutation relation, by using 
$\displaystyle{S_{\mu \alpha}(x)=x_{\nu}\gamma^{\nu}_{\alpha \beta}
J_{\mu \beta}(x)}$ as 
\bea
&&\langle 0|\{S_{\alpha},\not{\! \partial}_{\gamma \beta}
\bar{\chi}_{\gamma}(y)\}|0\rangle |_{x_0 =y_0}\nonumber\\
&=&
\int d^4x \ \delta(x_0 -y_0)x_{\mu}\gamma^{\mu}_{\alpha \delta}
\not{\! \partial}_{\gamma \beta}(y)
\langle 0|\{J_{0 \delta}(x),\bar{\chi}_
{\gamma}(y)\}|0\rangle \nonumber\\
&=&
4c\delta_{\alpha \beta}.\nonumber
\label{eqn:a17}
\eea
Thus if $Q$-supersymmetry is spontaneously broken, then $S$-supersymmetry 
is spontaneously broken as well. And the NG fermion for the broken 
$S$-supercharge is written by the derivative of the NG fermion 
for the broken $Q$-supercharge.

We now turn to spontaneous breaking of $D$=4 $N$=2 superconformal group 
down to $N$=1 super-Poincar{\'e} group, i.e. the breaking pattern B in Fig.1.
We derive the constraints imposed on NG particles as in the previous case. 
The commutation relations:
\bea
&&\{Q_{\alpha 1},S^{\beta 1}\}=(\sigma^{\mu \nu})_{\alpha}^{\ \beta}
M_{\mu \nu}-2i\delta_{\alpha}
^{\beta}D-4\delta_{\alpha}^{\beta}B_1^{\ 1}\nonumber\\
&&[Q_{\alpha 1},K_{\mu}]=\frac{1}{2} \sigma _{\mu \alpha \dot{\alpha}}
\bar{S}^{\dot{\alpha}}_1,\ 
[P_{\mu},K_{\nu}]=2i(g_{\mu \nu}D-M_{\mu \nu}) 
\label{eqn:a7}
\eea
and (\ref{eqn:a2},\ref{eqn:a5}) give constraints on $N$=1 NG multiplets. 
As before, Jacobi identities between (\ref{eqn:a7}) and NG particles lead to
\be
\psi_{\alpha 1}\sim [Q_{\alpha 1},\sigma +i\rho]\ ,\ 
\phi_{\mu}\sim \bar{\sigma}_{\mu}^{\dot{\alpha} \alpha}\{Q_{\alpha 1},
\bar{\psi}_{\dot{\alpha}1}\}\ ,\ \phi_{\mu}\sim \partial_{\mu}\sigma.
\label{eqn:a8}
\ee
These transformations (\ref{eqn:a8}) show that there exists another 
independent $N$=1 NG multiplet $(\sigma + i\rho)(x,\theta,\bar{\theta}) 
\equiv \phi \ (x,\theta,\bar{\theta})$ which consists of 
component fields $\phi$ and $\psi_{\alpha 1}$.
We put the chirality condition on $\phi \ (x,\theta,\bar{\theta})$.
Thus we conclude that $\phi \ (x,\theta,\bar{\theta})$ is the NG-background 
chiral superfield which contains a complex spin 0 field $\phi=\sigma+i\rho$, 
where $\sigma$ is the dilaton and $\rho$ is the axion, and spin $\frac{1}{2}$ 
field $\psi_{\alpha 1}$ is the dilatino. 

Now we turn to the coset construction for the nonlinear realization.
When $D$=4 $N$=2 superconformal group is spontaneously broken to $N$=1 
super-Poincar{\'e} group, the relevant left-invariant coset representative 
is given by 
\be
L(x,\ \theta,\ \bar{\theta})=T\ F\ U
\label{eqn:a19}
\ee
where
\bea
&&T=\exp(ix\cdot P+i\theta^A Q_A+i\bar{\theta}_A \bar{Q}^A)\ \ \ \ \ \ \ \ \ 
\ \ \ \theta^2 \equiv \chi\ ,\ \bar{\theta}_2 \equiv \bar{\chi}\nonumber\\
&&F=\exp(i\phi \cdot K+i\psi_A S^A+i\bar{\psi}^A \bar{S}_A)
\exp(i\sigma D+iv_1 B_1^{\ 1}+iv_2 B_2^{\ 2})\\
&&U=\exp(ivT+i\bar{v}\bar{T}).\nonumber
\label{eqn:a20}
\eea
Under the left multiplication of a group element $g\in G$, $L$ transforms as
\be
L \rightarrow gL=L'h \qquad (h\in H),
\ee
from which we obtain the transformation laws of the NG fields.
We calculate the Cartan differential 1-form as  
\bea
L^{-1}dL&=&i\mbox{D}x\cdot P+i\mbox{D}\theta^A Q_A+i\mbox{D}\bar{\theta}_A 
\bar{Q}^A
+i\mbox{D}\phi \cdot K+i\mbox{D}\psi_A S^A+i\mbox{D}\bar{\psi}^A \bar{S}_A
\nonumber\\
&&+i\mbox{D}\sigma D+i\mbox{D}v_1 B_1^{\ 1}
+i\mbox{D}v_2 B_2^{\ 2}+i\mbox{D}vT
+i\mbox{D}\bar{v}\bar{T}+\frac{1}{2}\omega_{\mu \nu}M^{\mu \nu}
\label{eqn:a21}
\eea
where
\bea
&&\mbox{D}x^{\mu}=\exp(-\sigma)[dx^{\mu}+id\theta^A\sigma^{\mu}
\bar{\theta}_A+id\bar{\theta}_A\bar{\sigma}^{\mu}\theta^A]\equiv 
\exp(-\sigma)dl^{\mu}\nonumber\\
&&\mbox{D}\theta^{\alpha C}=W_{A}^{\ B}U_B^{\ C}[d\theta^{\alpha A}-
i(\bar{\psi}^A \bar{\sigma}_{\mu})^{\alpha}dl^{\mu}]\nonumber\\
&&\mbox{D}\bar{\theta}_{\dot{\alpha}C}=(W^{\dagger})_{B}^{\ A}(U^{-1})
^{\ B}_C
[d\bar{\theta}_{\dot{\alpha}A}-i(\psi_A\sigma_{\mu})_{\dot{\alpha}}
dl^{\mu}]\nonumber\\
&&\mbox{D}\sigma =d\sigma -2[dl\cdot \phi -d\theta^A\psi_A -d\bar{\theta}_A
\bar{\psi}^A]\\
&&\mbox{D}\rho =d\rho -2[dl^{\mu}(\bar{\psi}^A\bar{\sigma}_{\mu}\psi_A)
+id\theta^A\psi_A -id\bar{\theta}_A\bar{\psi}^A]\nonumber\\
&&U_A^{\ B}\equiv \exp{i(v\tau +\bar{v}\bar{\tau})}_A^{\ B}
=
\bmat{cc}
\cos{\sqrt{v{\bar v}}} & i\sqrt{\frac{v}{\bar v}}\sin{\sqrt{v{\bar v}}} \\
i\sqrt{\frac{\bar v}{v}}\sin{\sqrt{v{\bar v}}} & \cos{\sqrt{v{\bar v}}} \\
\emat \nonumber\\
&&W_A^{\ B}\equiv \exp{[-\frac{1}{2} \{(\sigma -3i\rho)1 -iv_3 \tau_3
\}]_A^{\ B}}\ ,\ 
\rho\equiv \frac{1}{6}(v_1+v_2)\ ,\ v_3 \equiv v_1-v_2.\nonumber
\label{eqn:a22}
\eea
We take $Q_{2\alpha}$ to be the broken supercharge while keeping $Q_{1\alpha}$
unbroken. The NG field of the broken $Q_{2\alpha}$ is denoted by $\chi_\alpha$.
Here one should also note that we only consider the case where 
$SU(2)\times U(1)$ symmetrey is broken to $U(1)$. Namely, the three 
generators; $T={B_2}^1$, $\bar{T}={B_1}^2$ and 
$A=3({B_1}^1+{B_2}^2)$ are broken and $T_3=\frac{1}{2}({B_1}^1-{B_2}^2)$ 
remains unbroken. This is consitent with the relations obtained by the charge 
algebras.
Further, if $T_3$ were broken, we would need another real scalar as well as 
one fermionic superpartner, but there are no such particles in the present 
case. Therefore, it is reasonable to keep $T_3$ unbroken. This amounts to
take $v_3=0$ and $W_A^{\ B}$ is reduced to $\exp[-\frac{1}{2}(\sigma-3i\rho)]
\ \delta_A^{\ B}$.

Now let us proceed to the constraints to be imposed on particles in the 
framework of nonlinear realization in order to eliminate the 
unphysical degrees of freedom. Before doing this, we introduce the 
superspace coordinates $X^{A}=(x^{\mu}, 
\theta^{\alpha}, {\bar\theta}_{\dot\alpha})$ and the
supervielbein $E_M^{\ A}$ defined as 
\bea
\mbox{D}X^{A}=dX^ME_M^{\ A}\ ,\quad  \mbox{D}X^{A}\equiv 
( \mbox{D}x^{\mu},\ \mbox{D}\theta ^{\alpha},
\ \mbox{D}\bar{\theta}_{\dot{\alpha}})\ ,
\ dX^M\equiv ( dx^m,\ d\theta ^a,\ d\bar{\theta}_{\dot{a}}).
\label{eqn:a24}
\eea

In general the constraints should be invariant under the nonlinear 
transformations, and therefore its form is represented as 
$\mbox{D}_A\xi = \mbox{constant}$ for any NG particles $\xi$, where
\be
\mbox{D}_A\xi=\frac{\mbox{D}\xi}{\mbox{D}X^A}
=(E^{-1})_A^{\ M}\frac{\mbox{D}\xi}{dX^M} 
\ee
is the covariant derivative of NG field in the nonlinear realization.
This is because the covariant derivatives of NG particles transform 
linearly under the full group. In our present case we set the following 
constraints: 
\bea
&&\mbox{D}_\mu\chi^\alpha=\mbox{D}_{\dot\alpha}\chi^\alpha=
\mbox{D}_\alpha\chi^\alpha=0,\quad 
\mbox{D}_\mu{\bar\chi}_{\dot\alpha}=\mbox{D}_{\alpha}{\bar\chi}_{\dot\alpha}
=\mbox{D}^{\dot\alpha}{\bar\chi}_{\dot\alpha}=0, \nonumber\\ 
&&\mbox{D}_A\sigma =0\ (A=\mu,\alpha,{\dot\alpha});
\ \mbox{D}_\alpha \rho =0,\ \mbox{D}_{\dot\alpha} \rho =0.
\label{eqn:a23}
\eea
We will see that these constraints realize the relations 
(\ref{eqn:a6},\ref{eqn:a8}) and further impose the chirality conditions 
on $\chi$ as well as on $\phi \equiv \sigma+i\rho$.
Note that we do not require $\mbox{D}_\mu\rho=0$, because it would
lead to an over-constraint among the independent NG fields.
Moreover, other possible constraint like $\mbox{D}_A\psi_2=0$ does not match
the relations (\ref{eqn:a6},\ref{eqn:a8}).

Using the supervielbein matrix elements, we define covariant derivatives 
in the NG background as ($\partial_m=\partial/\partial x^m$, 
$\partial_a=\partial/\partial \theta^a$, 
$\partial^{\dot a}=\partial/\partial{\bar\theta}_{\dot a}$): 
\be 
{\cal D} _{\mu}\equiv {(e^{-1})_{\mu}}^{m}\partial_m \ 
,\ 
{\cal D} _{a}\equiv 
\partial_{a} - {e_{a}}^{\mu}
{\cal D}  _{\mu}\ ,\ 
\bar{{\cal D} }^{\dot{a}}
\equiv \partial^{\dot{a}} - e^{\dot{a} \mu}{\cal D} _{\mu}. 
\label{eqn:a25}
\ee
\bea
{e_m}^\mu&\equiv&\delta^{\ \mu}_{m}+i(\partial_m \chi \sigma^{\mu}
\bar{\chi}
+\partial_m\bar{\chi}\bar{\sigma}^{\mu}\chi),\quad {(e^{-1})_\mu}^m {e_m}^\nu
={\delta_\mu}^\nu \nonumber\\
e^{\ \mu}_{a}&\equiv&i(\sigma^{\mu}\bar{\theta})_a 
+i(\partial_a \chi \sigma^{\mu}\bar{\chi}
+\partial_a\bar{\chi}\bar{\sigma}^{\mu}\chi)\\
e^{\dot{a} \mu}&\equiv&i(\bar{\sigma}^{\mu}\theta)^{\dot{a}}
+i(\partial^{\dot{a}} \chi \sigma^{\mu}\bar{\chi}
+\partial^{\dot{a}}\bar{\chi}\bar{\sigma}^{\mu}\chi)\nonumber
\eea
which coincide with those introduced in \cite{BG2}.  Now we solve 
(\ref{eqn:a23}) and find
\bea
&&\bar{{\cal D}}^{\dot{\beta}}\chi^{\alpha}=0\ ,\ 
\bar{{\cal D}}^{\dot{\beta}}\phi =0 \label{chirality}\\
&&{\bar\psi}_{\dot\alpha}^1 =-\frac{1}{4}\ ({\bar Y}^{-1})_{\dot\alpha}^
{\dot\beta}[\ \bar{\cal D}_{\dot\beta}\bar{\phi}+i{\cal D}_\mu(\chi\sigma^\mu)
_{\dot\gamma}\bar{\cal D}_{\dot\beta}\bar{\chi}^{\dot\gamma}],\quad 
{\bar Y}_{\dot\beta}^{\dot\alpha} \equiv 
\delta_{\dot\beta}^{\dot\alpha}+\frac{U_1^{\ 2}}{U_2^{\ 2}}
\bar{\cal D}_{\dot\beta}\bar{\chi}^{\dot\alpha}
\label{dilatino}\\
&&
{\bar\psi}_{\dot\alpha}^2 =-\frac{1}{4}\ ({\bar Y}^{-1})_{\dot\alpha}^
{\dot\beta}
[\ i{\cal D}_\mu(\chi\sigma^\mu)_{\dot\beta}
-\frac{U_1^{\ 2}}{U_2^{\ 2}}\bar{\cal D}_{\dot\beta}\bar{\phi}] 
\label{s2fermi}\\
&&i\sqrt{\frac{v}{\bar{v}}}\tan{2\sqrt{v \bar{v}}}=
({\cal D}\chi +\frac{1}{2}\bar{\cal D}\bar{\chi}
{\cal D}_{\alpha}\chi_{\beta}{\cal D}^{\alpha}{\chi}^{\beta})\ 
(1-\frac{1}{4}{\cal D}_{\gamma}\chi_{\delta}{\cal D}^{\gamma}{\chi}^{\delta}
\bar{\cal D}_{\dot \alpha}{\bar\chi}_{\dot \beta}\bar{\cal D}^{\dot \alpha}
{\bar\chi}^{\dot \beta})
\label{su2}\\
&&\phi_{\mu}=\frac{1}{2}{\cal D}_{\mu}\sigma +
({\cal D}_{\mu}\chi)\psi_2+
({\cal D}_{\mu}\bar{\chi})\bar{\psi}^2 \label{conboost}
\eea
Note that the conjugated relations also hold for (\ref{chirality})
-(\ref{su2}). The two equations of (\ref{chirality}) are the chirality 
conditions for $\chi^\alpha$ and $\phi=\sigma+i\rho$.  
From (\ref{dilatino}), the superfield of the dilatino $\psi_1$, the 
superpartner of the dilaton, is given by the other superfields. 
The equation (\ref{s2fermi}) gives the superfield $\psi_2$ as derivatives 
of ${\bar\chi}$ and $\phi$, in accordance with the statement we made before 
by using the charge algebras and the spectral representation. 
The equation (\ref{su2}) shows that the NG field of the broken
SU(2) generators, $T$ and ${\bar T}$, is given by the superfield $\chi$.
The last equation (\ref{conboost}) indicates that $\phi_\mu$ is not
an independent NG degree of freedom but is given by other NG fields.

The explicit expression for the supervielbein in the full superspace 
is given by
\be
{E_M}^A=
\bmat{ccc}
{A_m}^\mu & {D_m}^\alpha & G_{m{\dot\alpha}}\\
{B_a}^\mu & {E_a}^\alpha & H_{a{\dot\alpha}} \\
C^{{\dot a}\mu} & F^{{\dot a}\alpha} & J^{\dot a}_{\ \dot\alpha} \\
\emat
\label{sdet}
\ee
where
\bea
&&{A_m}^\mu=e^{-\sigma}{e_m}^\mu, 
\quad {B_a}^\mu=e^{-\sigma}{e_a}^\mu, \quad C^{{\dot a}\mu}=
e^{-\sigma}e^{{\dot a}\mu}
\nonumber\\
&&{D_m}^\alpha=e^{-\frac{1}{2}(\sigma-3i\rho)}\left[\partial_m\chi^\alpha U_2^1
-i{e_m}^\mu(\bar{\psi}^A\bar{\sigma}_\mu)^\alpha U_A^1\right]
\nonumber\\
&&{E_a}^\alpha=e^{-\frac{1}{2}(\sigma-3i\rho)}\left[U_1^1\delta_a^\alpha+U_2^1
\partial_a\chi^\alpha-ie_a^\mu(\bar{\psi}^A\bar{\sigma}_\mu)^\alpha U_A^1
\right]
\nonumber\\
&&F^{{\dot a}\alpha}=e^{-\frac{1}{2}(\sigma-3i\rho)}
\left[U_2^1\partial^{\dot a}\chi^\alpha-ie^{{\dot a}\mu}(\bar{\psi}^A\bar
{\sigma}_\mu)U_A^1\right]
\nonumber\\
&&G_{m{\dot\alpha}}=e^{-\frac{1}{2}(\sigma+3i\rho)}
\left[\partial_m\bar{\chi}_{\dot\alpha}(U^{-1})_1^2-ie_m^\mu
(\psi_A\sigma_\mu)_{\dot\alpha}(U^{-1})_1^A\right]
\nonumber\\
&&H_{a{\dot\alpha}}=e^{-\frac{1}{2}(\sigma+3i\rho)}
\left[(U^{-1})_1^2\partial_a\bar{\chi}_{\dot\alpha}-ie_a^\mu
(\psi_A\sigma_\mu)_{\dot\alpha}(U^{-1})_1^A\right]
\nonumber\\
&&{J^{\dot a}}_{\dot\alpha}=e^{-\frac{1}{2}(\sigma+3i\rho)}
\left[(U^{-1})_1^1\delta_{\dot\alpha}^{\dot a}+(U^{-1})_1^2\partial^{\dot a}
\bar{\chi}_{\dot\alpha}-ie^{{\dot a}\mu}(\psi_A\sigma_\mu)_{\dot\alpha}
(U^{-1})_1^A \right]
\eea
From these result, we can compute the superdeterminant of $E_M^{\ A}$ as
\bea
\hspace{-0.5cm}\mbox{sdet}\ (E_M^{\ A})&=&\det\ ({A_m}^\mu)\cdot{\det}^{-1}
({\hat D}-{\hat B}A^{-1}{\hat C}) \nonumber\\
&=&e^{-2\sigma}\det\ ({e_m}^\mu)
\left|1-\frac{i}{2}{\cal D}\chi\sqrt{\frac{\bar v}{v}}\tan{\sqrt{v{\bar v}}}
\right|^{-2}
\eea
where we have denoted the submatrices as
\bea
&&E_M^{\ A}=
\bmat{cc}
{A_m}^\mu & {\hat C} \\
{\hat B} & {\hat D}
\emat, \qquad
{\hat B}=
\bmat{c}
{B_a}^\mu \\ C^{{\dot a}\mu}
\emat \nonumber\\
&&
{\hat C}=
\bmat{cc}
{D_m}^\alpha & G_{m{\dot\alpha}}
\emat,
\qquad 
{\hat D}=
\bmat{cc}
{E_a}^\alpha & H_{a{\dot\alpha}} \\
F^{{\dot a}\alpha} & {J^{\dot a}}_{\dot\alpha}
\emat.
\eea
The inverse of the supervielbein, $(E^{-1})_A^{\ M}$ is computed in a similar
way.

The simplest invariant action for the dilaton multiplet is obtained from 
the above superdeterminant, by forming the invariant phase volume.
\bea
&&\int d^4 xd^2 \theta d^2\bar{\theta}\ \mbox{sdet}E_M^{\ A}\nonumber\\
&=&\int d^4 x
{\cal L}_D + \cdots
\label{eqn:a27}
\eea
where the nonlinear lagrangian ${\cal L}_D$ turns out to be
\bea
&&{\cal L}_D=
e^{-\varphi-\varphi^*}
\left[-\partial^m\varphi^*\partial_m\varphi
+\frac{i}{2}\partial_m\bar{\psi}\bar{\sigma}^m\psi
-\frac{i}{2}\bar{\psi}\bar{\sigma}^m\partial_m\psi\right.
\qquad\qquad\qquad\qquad \nonumber\\
&&\qquad\qquad\qquad-
\left.\frac{i}{2}(\partial_m\varphi-\partial_m\varphi^*)\bar{\psi}
\bar{\sigma}^m\psi+(F^*+\frac{1}{2}\bar{\psi}^2)
(F+\frac{1}{2}\psi^2)\right]
\label{dilalag}
\eea
where we denote the first, second and auxiliary components of the superfield 
$\phi$ by $\varphi$, $\psi$ and $F$, respectively. The $\psi$ is, at the
same time, the first component of the superfield $\psi_1$. 
This is the same effective lagrangian for the sponatneous breaking of N=1
superconformal to N=1 super-Poincar{\'e} symmetry corresponding to the pattern
D in Fig.1 and discussed in ref.\cite{KU}.  
 
Now we introduce the chiral superspace in the NG background \cite{BG2}
, $(x_L^{\ m}, \theta^\alpha)$, as
\be
x_L^{\ m}=x^m-i\theta\sigma^m\bar{\theta}-i\chi\sigma^m\bar{\chi}
\ee

The supervielbein ${E_L}_M^{\ A}$ for the chiral superspace in NG 
background is
\be
\bmat{cc} 
Dx^\mu, & D\theta^\alpha 
\emat
=
\bmat{cc} 
{dx_L}^m, & d\theta^a 
\emat
\bmat{cc}
{A_m}^\mu & {C_m}^\alpha \\
{B_a}^\mu & {D_a}^\alpha 
\emat
\equiv
\bmat{cc} 
{dx_L}^m, & d\theta^a 
\emat
E_L
\label{chiraldet}
\ee
where
\bea
&&{A_m}^\mu=e^{-\sigma}e_{Lm}^\mu, \quad e_{Lm}^\mu=\delta_m^\mu+
2i\partial_m^L\chi\sigma^\mu{\bar\chi} \nonumber\\
&& B_a^\mu= e^{-\sigma}e_{La}^\mu, \quad e_{La}^\mu=2i(\sigma^\mu{\bar\theta}
)_a+2i\partial_a\chi\sigma^\mu{\bar\chi} \nonumber\\
&&C_m^\alpha=e^{\frac{1}{2}(\sigma-3i\rho)}\left(\partial_m^L\chi^\alpha
U_2^1-ie_{Lm}^\mu(\bar{\psi}^A\bar{\sigma}_\mu)^\alpha U_A^1 \right)
\nonumber\\
&&D_a^\alpha=e^{\frac{1}{2}(\sigma-3i\rho)}\left(\delta_a^\alpha U_1^1+
\partial_a\chi^\alpha U_2^1-ie_{La}^\mu(\bar{\psi}^A\bar{\sigma}_\mu)^\alpha 
U_A^1 \right)
\eea

The chiral superdeterminant is given by
\be
\mbox{sdet} E_L= \det A\cdot {\det}^{-1}(D-BA^{-1}C)
=e^{-3\sigma-3i\rho}\det (e_{Lm}^\mu)
[1-\frac{i}{2}\sqrt{\frac{\bar v}{v}}\tan{\sqrt{v{\bar v}}}
{\cal D}\chi]^{-1}.
\ee

The invariant nonlinear lagarangian has the following form
for the chiral superspace:
\be
\int d^4 x_L d^2\theta\  \mbox{sdet} E_L \ f( D_A\xi, \Phi) +\mbox{h.c.}
\label{chilag}
\ee
where $\xi$ is any NG field and $\Phi$ is a spectator field which transforms
linearly in the nonlinear realization.

Now we construct the invariant lagrangian for the vector multiplet of 
NG fermion for the broken supercharge $Q_2$. 

Under the supersymmetry transformation of $Q_2$: 
$g=\exp{i(\eta Q_2+{\bar\eta}{\bar Q}_2)}$,
\bea
&&x^m \rightarrow x'^m=x^m+i(\eta\sigma^m{\bar\chi}-\chi\sigma^m{\bar\eta})
\nonumber\\
&& \theta^\alpha \rightarrow \theta'^\alpha=\theta^\alpha,
\quad {\bar\theta}_{\dot\alpha} \rightarrow {\bar\theta}'_{\dot\alpha}
={\bar\theta}_{\dot\alpha} \label{trans}
\eea
and the NG fermion varies as 
\be
\chi'^\alpha(x',\theta',{\bar\theta}')=\chi^\alpha(x,\theta,{\bar\theta})
+\eta^\alpha, \quad
{\bar\chi}'_{\dot\alpha}(x',\theta',{\bar\theta}')
={\bar\chi}_{\dot\alpha}(x,\theta,{\bar\theta})+{\bar\eta}_{\dot\alpha}
\ee
Note that the NG fermion field $\chi_\alpha$ starts with $W_\alpha=
i{\bar D}^2D_\alpha V$, 
\be
\chi_\alpha=W_\alpha+\frac{1}{4}{\bar D}^2({\bar W}^2)W_\alpha
-iW\sigma^\mu{\bar W}\partial_\mu W_\alpha +{\cal O}(W^5).
\ee
where $D_\alpha=
\partial_\alpha-i(\sigma^\mu{\bar\theta})_\alpha\partial_\mu$ and
${\bar D}^{\dot\alpha}=\partial^{\dot\alpha}-i(\sigma^\mu{\theta})^
{\dot\alpha}\partial_\mu$ are the ordinary supercovariant derivative and 
$V(x,\theta,{\bar\theta})$ is the vector superfield \cite{BG2}.

The invariant effective lagrangian for the vector multiplet can be
obtained by the standard prescription (\ref{chilag}) in the nonlinear 
realization. We take the function $f$ to be
\be
f(\mbox{D}_A\xi)=\frac{1}{3}
\left[1+\frac{1}{2}\mbox{D}_{({\dot\alpha}}{\bar\chi}_{\dot\beta)}
\mbox{D}^{({\dot\alpha}}{\bar\chi}^{\dot\beta)} \right]
\ee
where
\be
\mbox{D}_{({\dot\alpha}}{\bar\chi}_{\dot\beta)}
=\frac{1}{2}(\mbox{D}_{{\dot\alpha}}{\bar\chi}_{\dot\beta}+
\mbox{D}_{{\dot\beta}}{\bar\chi}_{\dot\alpha})
\ee
is the symmetric part and is equal to
\be
\mbox{D}_{({\dot\alpha}}{\bar\chi}_{\dot\beta)}
=\frac{(U^{-1})_2^{\ 2}}{\bar\Delta}\bar{\cal D}_
{({\dot\alpha}}{\bar\chi}_{\dot\beta)}=
\bar{\cal D}_{({\dot\alpha}}{\bar\chi}_{\dot\beta)}+\cdots
\ee
with
\be
{\bar\Delta}=1+\frac{i}{2}\sqrt{\frac{v}{\bar v}}\tan{\sqrt{v{\bar v}}}
\bar{D}\bar{\chi}
\ee

The effective lagrangian for the vector multiplet (or NG-Maxwell multiplet)
arises from the interplay of the D-term (\ref{eqn:a27}) and the F-term given
as follows,
\bea
&&\int d^4x_Ld^2\theta \ \frac{1}{3} e^{-3\phi}\det (e_{Lm}^\mu)
[1-\frac{i}{2}\sqrt{\frac{\bar v}{v}}\tan{\sqrt{v{\bar v}}}{\cal D}\chi]^{-1}
[1+\frac{1}{2}\mbox{D}_{({\dot\alpha}}{\bar\chi}_{{\dot\beta})}
\mbox{D}^{({\dot\alpha}}{\bar\chi}^{{\dot\beta})}]+\mbox{h.c.}\nonumber\\
&&=\int d^4 x_L d^2\theta e^{-3\phi}\ (1+2i\partial_m^L\chi\sigma^m{\bar\chi}
+\cdots)(1-\frac{1}{4}{\cal D}\chi\bar{\cal D}\bar{\chi}+\cdots)
\nonumber\\
&&\qquad\qquad\qquad\times(1-\frac{1}{4}(\bar{\cal D}\bar{\chi})^2+
\frac{1}{2}\bar{\cal D}_{\dot \alpha}\bar{\chi}_{\dot \beta}
\bar{\cal D}^{\dot \alpha}\bar{\chi}^{\dot \beta}) + \mbox{h.c.}\nonumber\\
&&=\int d^4x_L {\cal L}_{NGM}
\eea
where the following relation has been used:
\be
\bar{\cal D}_{({\dot\alpha}}{\bar\chi}_{{\dot\beta})}
\bar{\cal D}^{({\dot\alpha}}{\bar\chi}^{{\dot\beta})}
=\bar{\cal D}_{{\dot\alpha}}{\bar\chi}_{{\dot\beta}}
\bar{\cal D}^{{\dot\alpha}}{\bar\chi}^{{\dot\beta}}
-\frac{1}{2}(\bar{\cal D}\bar{\chi})^2. 
\ee
To the order of ${\bar\chi}^2={\bar W}^2+\cdots$ we get
\be
{\cal L}_{NGM}=\int d^2\theta \ e^{-3\phi}\ \frac{1}{3} (
1-\frac{1}{4}{\bar D}^2 (\bar{W}^2)+\cdots) +\mbox{h.c.}
\ee
where the integrand turns out to be chiral to this order. In this sense
the choice of $f(\mbox{D}_A\xi)$ is unique to the order of ${\bar\chi}^2$.
Note that we have used the Bianchi identity, $DW=-{\bar D}{\bar W}$.
The total lagrangian
\be
{\cal L}={\cal L}_D+{\cal L}_{NGM}
\ee
provides the kinetic term for the vector multiplet by using the equation
of motion for $F$ and $F^{*}$. Hence we get
\be
-\int d^4x e^{-4\sigma}
\left\{1+\frac{1}{4}F_{mn}F^{mn}-\frac{i}{2}\partial_m{\bar\lambda}\sigma^m
\lambda+\frac{i}{2}{\bar\lambda}{\bar\sigma}^m\partial_m\lambda
-\frac{1}{2}D^2+{\cal O}(F_{mn}^4)\right\}
\label{kinetic}
\ee
This action is invariant under the nonlinear scale transformation: $x^\mu
\rightarrow e^\kappa x^\mu$, where the dilaton transforms as 
$\sigma \rightarrow \sigma +\kappa$ and the gauge field strength $F_{mn}$
as well as the auxiliary field $D$ have vanishing scale dimensions and 
$\lambda$ has a scale dimension 1/2. This is because the superfield $\chi$ 
has a scale dimension 1/2, as can be seen from the coset construction. 

Here we should note the connection with the breaking pattern C,
which was studied in ref. \cite{KLU}.  In that case
there appear a $N=2$ multiplet consisting of dilaton $\sigma$, axion $\rho$,
dilatinos $\psi_i$, ${\bar\psi}^i (i=1,2)$ and vector gauge field $A_\mu$.
The effective lagrangian in that case is similar to the sum of 
(\ref{dilalag}) and (\ref{kinetic}).
If the $N=2$ supersymmetry breaks down to $N=1$, this multiplet splits into
two $N=1$ multiplets; the vector multiplet and the chiral multiplet.

Finally, a comment on the Born-Infeld action \cite{BI,CF} is in order.
It is not straightforward to extend the Born-Infeld action  
for the partially broken $N=2$ super-Poincar{\'e} symmetry discussed
by Bagger and Galperin \cite{BG2} to the present case, where the dilaton 
multiplet is coupled to the NG-Maxwell multiplet. It might be possible to
derive the Born-Infeld action by summing up the higher order powers of
$F_{mn}$ in (\ref{kinetic}), which is now under investigation.

%%%%%%%%%%%%%%%%%%%% Figure caption %%%%%%%%%%%%%%%%%%
\newpage
\vspace{3cm}
\noindent
{\large Figure Caption}
\baselineskip 16pt

\vspace{0.5cm}
\noindent
Fig.1 \quad 
Various patterns for spontaneous breaking of $N=2$ superconformal symmetry 
\setlength{\unitlength}{1mm}
\begin{picture}(100,100)(-50,-40)
\put(0,40){\framebox(40,10){$N$=2 superconformal}}
\put(-40,0){\framebox(40,10){$N$=1 superconformal}}
\put(40,0){\framebox(40,10){$N$=2 super-Poincar{\'e}}}
\put(0,-40){\framebox(40,10){$N$=1 super-Poincar{\'e}}}
\put(10,40){\vector(-1,-1){30}}
\put(30,40){\vector(1,-1){30}}
\put(-20,0){\vector(1,-1){29.5}}
\put(60,0){\vector(-1,-1){29.5}}
\put(20,40){\vector(0,-1){69.5}}
%\put(-20,10){\line(1,1){25}}
%
\put(-15,20){\makebox(10,10){A}}
\put(20,0){\makebox(10,10){B}}
\put(45,20){\makebox(10,10){C}}
\put(-15,-20){\makebox(10,10){D}}
\put(45,-20){\makebox(10,10){E}}
\end{picture}


\newpage
\baselineskip 17pt
\begin{thebibliography}{99}

\bibitem{HLP}
     J.~Hughes, J.~Liu and J.~Polchinski, {\sl Phys.~Lett.} 
     {\bf B180} (1986) 370;\\
     J.~Hughes and J.~Polchinski, {\sl Nucl.~Phys.} {\bf B278} (1986) 147.

\bibitem{APT}
     I.~Antoniadis, H.~Partuche and T.~R.~Taylor, {\sl Phys.~Lett.} 
     {\bf B372} (1996) 83; \\
     I.~Antoniadis and T.~R.~Taylor, CPTH-PC445.0496, hep-th/9604062(1996);\\
     H.~Partouche and B.~Pioline, CPTH-PC496.0297, hep-th/9702115 (1997).

\bibitem{FGP}
     S.~Ferrara, L.~Girardello and M.~Poratti {\sl Phys.~Lett.} {\bf B376} 
     (1996) 275; \\
     M.~Poratti,hep-th/9609073 (1996);\\
     P.~Fr{\'e}, L.~Girardello, I.~Pensando and M.~Trigiante, hep-th/9611188
     (1996).

\bibitem{BG1}
     J.~Bagger and A.~Galperin, {\sl Phys.~Lett.} {\bf B336} (1994) 25.

\bibitem{BG2}
     J.~Bagger and A.~Galperin, {\sl Phys.~Rev.} {\bf D55} (1997) 1091.
      
\bibitem{BW}
     J.~Bagger and J.~Wess, {\sl Phys.~Lett.} {\bf 138B} (1984) 105.

\bibitem{SW}
     S.~Samuel and J.~Wess, {\sl Nucl.~Phys.} {\bf B221} (1983) 153.

\bibitem{B}
     J.~Bagger, {\sl Nucl.~Phys.}~{\bf B}~(Proc.~Suppl.){\bf 52A} (1997) 362.

\bibitem{V}
     D.~V.~Volkov, {\sl Sov.~J.~Particles and Nuclei} {\bf 4} (1973) 3.

\bibitem{O}
     V.~I.~Ogievetsky, in Proc. of 10th Winter School of Theoretical
     Physics in Karpacz, Wroclaw (1974) vol.1, p.117.

\bibitem{UZ}
     T.~Uematsu and C.~K.~Zachos, {\sl Nucl.~Phys.} {\bf B201} (1982) 250.

\bibitem{CWZ}
     S.~Coleman, J.~Wess and B.~Zumino, {\sl Phys.~Rev.} {\bf 177} (1969)
     2239; \\
     C.~Callan, S.~Coleman, J.~Wess and B.~Zumino, {\sl Phys.~Rev.} 
     {\bf 177} (1969) 2247.

\bibitem{Fayet}
     P.~Fayet, {\sl Nucl.~Phys.} {\bf B149} (1979) 137.

\bibitem{Hig}
     K.~Higashijima, Proc. XX Int. Colloquium on Group Theoretical
Methods in Physics, July 4-9, 1994 Toyonaka, Japan (World Scientific, 1995),
pp. 223-228.

\bibitem{KU}
     K.~Kobayashi and T.~Uematsu, {\sl Nucl.~Phys.} {\bf B263} (1986) 309.

\bibitem{KLU}
     K.~Kobayashi, K.~H.~Lee and T.~Uematsu, {\sl Nucl.~Phys.} 
    {\bf B309} (1988) 669.
 
\bibitem{BI}
     M.~Born and L.~Infeld, {\sl Proc. Roy. Soc. London} {\bf A144} (1934) 425.
              
\bibitem{CF}
     S.~Cecotti and S.~Ferrara, {\sl Phys.~Lett.} {\bf B187} (1986) 335.


\end{thebibliography}
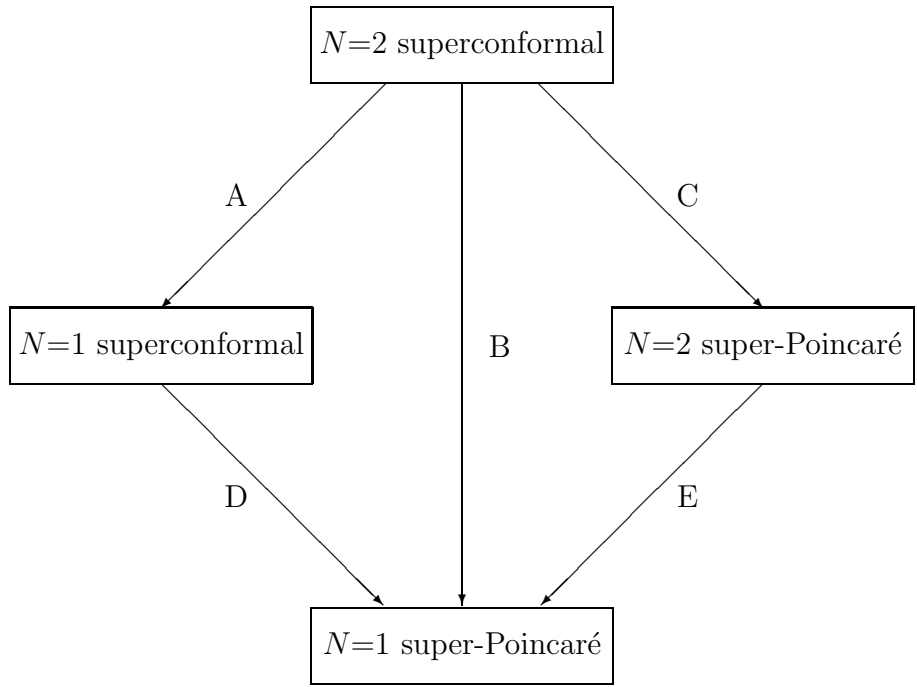
\end{document}